\documentclass[twocolumn]{aastex63}

\shorttitle{Disk Illumination and Jet Variability of the Herbig Ae Star HD 163296}
\shortauthors{Rich et al.}

\newcommand{\Msun}{M_\odot}
\newcommand{\Rsun}{R_\odot}
\newcommand{\Lsun}{L_\odot}
\accepted{August 24, 2020}
\submitjournal{AJ}

\shortauthors{Rich et al.}

\begin{document}

\title{Disk Illumination and Jet Variability of the Herbig Ae Star HD 163296 Using Multi-Epoch HST/STIS Optical, Near-IR, and Radio Imagery and Spectroscopy}

\correspondingauthor{Evan A. Rich}
\email{earich@umich.edu}

\author{Evan A. Rich}
\affiliation{Homer L. Dodge Department of Physics
University of Oklahoma 
Norman, OK 73071, USA}
\affiliation{Astronomy Department 
University of Michigan 
Ann Arbor, MI 48109, USA}

\author{John P. Wisniewski}
\affiliation{Homer L. Dodge Department of Physics
University of Oklahoma 
Norman, OK 73071, USA}

\author{Michael L. Sitko}
\affiliation{Department of Physics 
University of Cincinnati  
Cincinnati, OH 45221, USA}
\affiliation{Space Science Institute 
475 Walnut Street, Suite 205 
Boulder, CO 80301, USA}

\author{Carol A. Grady}
\affiliation{Exoplanets and Stellar Astrophysics Laboratory Code 667, Goddard Space Flight Center 
Greenbelt, MD 20771, USA}
\affiliation{Eureka Scientific 
2452 Delmer, Suite 100 
Oakland CA 96002, USA}
\affiliation{Goddard Center for Astrobiology}

\author{John J. Tobin}
\affiliation{National Radio Astronomy Observatory, 520 Edgemont Rd., Charlottesville, VA, 22903}

\author{Misato Fukagawa}
\affiliation{National Astronomical Observatory of Japan, 2-21-1, Osawa, Mitaka, Tokyo, 181-8588, Japan}

\begin{abstract}

We present two new epochs of Hubble Space Telescope/Space Telescope Imaging Spectrograph coronagraphic imaging, along with multi-epoch optical, near-IR, and radio monitoring, of the HD 163296 system. We find ansae features identified in earlier epoch HST imagery are a 4th ring, that resides at a semi-major axis distance of 3$\farcs$25 (330 au). We determine the scale height of the dust is 64 au at a radial distance of 330 au. We observe surface brightness variations in the 4th ring on $<$3 month timescales, including large-scale, azimuthally asymmetric changes. This variability resembles earlier studies of the innermost disk ring ($0 \farcs 66$, 67 au), suggesting a common origin. We find no evidence for the ejection of new HH-knots predicted to occur in 2018. Moreover, our non-detection of older HH-knots indicate the knots could be experiencing less shock-heating. We also detect one clear dipper event in our optical light curve from 2018. Using the time-scale and spatial extent of disk illumination changes we observe, we estimate the source of this shadowing resides within 0.5 au from the star, must extend at least 0.08 au above the midplane of the disk, and has an azimuthal extent of 0.26 au.  We estimate the source of the dipper event reaches a scale height of 0.37 au above the midplane at 0.41 au, and has an azimuthal extent of 0.3 au.  We suggest these similarities could indicate the same (or similar) mechanisms are responsible for producing both dippers and variable ring illumination in the system.

\end{abstract}


\section{Introduction}

Protoplanetary disks are dust and gas disks found around young stars that serve as the birthplace of planets.  Resolved imagery of Herbig Ae/Be protoplanetary disk systems has revealed a wealth of structure such as spiral arms \citep{hashimoto2011,grady2013,muto2012}, flat settled disks \citep{meeus2001}, and rings and gaps (e.g., TW Hydrae; \citealt{andrews2016}). These features are of interest as some may be indicators of Jovian-mass planets forming within the protoplanetary disk \citep{Quanz2013,Currie2015a}.  For example, the recent discoveries of PDS 70b and c and their circumplanetary disks \citep{keppler2018, muller2018, haffert2019} are located within a ring of the protoplanetary disk.

Young intermediate-mass stars, Herbig Ae/Be \citep{herbig1960}, are the more massive analogs to T-Tauri stars. They not only host protoplanetary disks but also commonly exhibit collimated bi-polar jets \citep{herbig1950,wassell2006,ellerbroek2014,bally2016}.  HD 163296 is a young ($5.1^{+0.3}_{-0.8}$ Myr old \citealt{montesinos2009} to $9 \pm 0.5$ Myr old \citealt{pikhartova2019}) Herbig Ae system located at a distance of 101.5 $\pm$ 1.2 $pc$ \citep{gaia2016,gaia2018}. 
The disk has been spatially resolved by ground- and space-based observing platforms at a multitude of wavelengths, including: optical (HST/STIS: \citealt{grady2000}, HST/ACS \citealt{wisniewski2008}), near-infrared (IR) (Subaru/HiCIAO: \citealt{rich2019}, Subaru/CHARIS: \citealt{rich2019}, VLT/NACO: \citealt{garufi2014,garufi2017}, Gemini/GPI: \citealt{monnier2017}, VLT/SPHERE: \citealt{muro2018}, Subaru/CIAO: \citealt{fukagawa2010}, Keck/NIRC2: \citealt{guidi2018}, and CHARA: \citealt{tannirkulam2008,setterholm2018}), and radio wavelengths (VLA: \citealt{guidi2016}, ALMA: \citealt{guidi2016,isella2016,isella2018,dent2019}).

\begin{figure}
\centering
\includegraphics[width=\columnwidth]{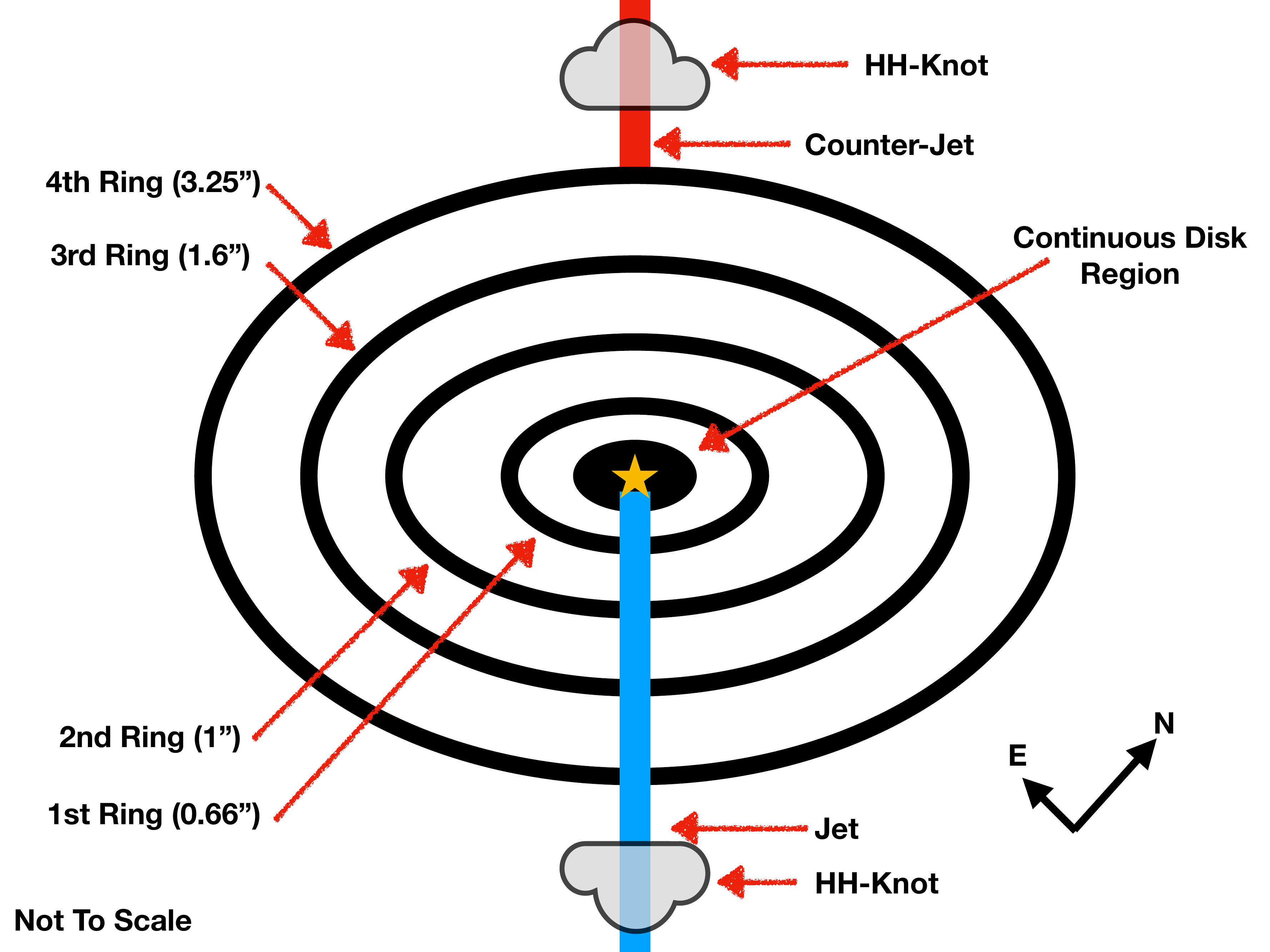}
\caption{A diagram of the HD 163296 system structure showing the known rings (black ellipses), and the orientation of the jet and counter-jet (shown in red and blue respectively). Generic HH-knots (grey) are also along the jet axis. Note that the diagram is not to scale, and the major axis is rotated to be horizontal. A compass in the lower left is provided for cardinal direction orientation.}\label{fig:diagram}
\end{figure}

These imaging studies have revealed that HD 163296 has a highly structured disk, as diagramed in Figure \ref{fig:diagram}. Closest to the star, there is a small ring at $0 \farcs 14$ (15 au) inside a continuous disk region \citep{isella2018}. Outside of the continuous disk region, there are three consecutive dust rings at $0 \farcs 66$, $1 \farcs 0$, and $1 \farcs 6$ (67 au, 102 au, and 160 au; \citealt{dent2019}) with gaps in between each ring.
The first dust ring at $0 \farcs 66$ has been detected with both radio/sub-mm \citep{guidi2016,isella2016,isella2018,dent2019} and near-IR observations \citep{garufi2014,garufi2017,monnier2017,muro2018,rich2019}. 
The dust disk extends to at least $4 \farcs 4$ (447 au) \citep{wisniewski2008} and two ansae (broken rings) have been detected at $2 \farcs 9$ (294 au) SE of the star and $3 \farcs 2$ (325 au) NW of the star \citep{grady2000}. 
We note that we choose to refer to these features as ansae to keep labeling consistent between published works (e.g. \citealt{grady2000,wisniewski2008}), however later in this paper, we do reinterpret these two ansae as a 4th broken ringed structure. Finally, the disk has exhibited time-dependent flux variability \citep{wisniewski2008,rich2019}, on timescales of $<$ 4 years.

The HD 163296 system hosts three planetary candidates. Modeling of ALMA gas emission data suggest Jovian-mass planets at $0 \farcs 82$ and $1 \farcs 35$ (83 and 137 au; \citealt{teague2018}) and a single Jovian mass planet on an even wider orbit at $2 \farcs 56$ \citep[260 au][]{pinte2018}. A fourth planet candidate had been suggested from Keck/NIRC2 thermal infrared imaging \citep[7 $M_{\rm J}$;][]{guidi2018}
just outside the first dust ring, but follow-up observations have failed to confirm this candidate   \citep{rich2019,mesa2019}.

HD 163296 also has had an active bi-polar jet HH409 first discovered through space-based coronagraphic images and verified with long-slit spectroscopy \citep{grady2000, devine2000}. 
Since then, several studies have detected additional knots and further characterized properties of the jet \citep{wassell2006,gunther2013}. 
In particular, \citet{ellerbroek2014} measured the proper motion and radial velocities of 11 HH-knots associated with HD 163296 and predicted that the system ejects HH-knots with a regular period of 16 $\pm$ 0.7 yr. 
\citet{sitko2008} and \citet{ellerbroek2014} have presented tentative evidence of optical flux dimming and IR flux increases on similar timescales. \citet{ellerbroek2014} also identified additional episodes of the host star dimming that could be similar to "dipper" events that have previously been seen in other protoplanetary systems \citep{cody2014,pinilla2018}. With the last HH-knot thought to have been ejected in 2002, the next HH-knot launch would be predicted to occur in 2018.

In this paper, we present new observations of the HD 163296 disk and jet that coincide with the expected next launch of a HH-knot \citep{ellerbroek2014}.  Our new observations include two epochs of Hubble Space Telescope (HST) coronagraphic imaging of HD 163296's protoplanetary disk with the HST Space Telescope Imager and Spectrograph (STIS), new multi-wavelength monitoring of the star from 2016-2018, new Lyman $\alpha$ observations of the jet with HST/STIS, as well as observations with the National Science Foundation's Karl G. Jansky Very Large Array (VLA) of the system to search for the launch of the predicted HH-knot. We describe the acquisition and reduction of these new data in Section \ref{sec:observations}. Next we analyze the HST/STIS data in Section \ref{sec:analysis} and the multi-wavelength observations in Section \ref{sec:analysis_mult}. Finally, we discuss our results in Section \ref{sec:discussion} and our conclusions in Section \ref{sec:conclusion}.

\section{Observations and Data Reduction}\label{sec:observations}
\subsection{AAVSO Observations}

\begin{figure*}
\centering
\includegraphics[width=\textwidth]{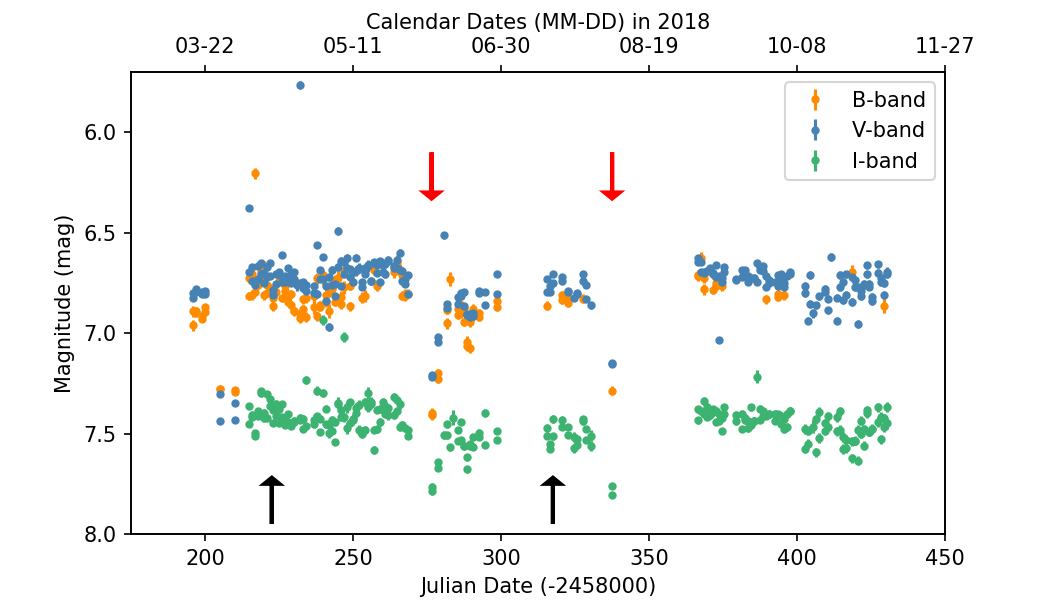}
\caption{Optical light curve for HD 163296 in 2018 in the B-,V-, and I-, bandpasses. The two black arrows denote epochs at which HST/STIS data were obtained. We note two potential dipper events in the light curve, depicted by red arrows, that happend on 2018 June 7 and 2018 Aug 7. \label{fig:AAVOS_lightcurve}}
\end{figure*}

Optical observations of HD 163296 were taken from 2018 March 18 to 2018 November 7 in the B-, V-, and I- bands with an observing cadence of $\sim$1 observation per day \citep{kafka2019}. Data were obtained through a request via AAVSO and taken with observers associated with Vereniging Voor Sterrenkunde, Werkgroep Veranderlijke Sterren (Belgium). 
The standard stars UCAC4 341-118672 and UCAC4 340-118109 were observed with the science target and were utilized in the reduction. The data were reduced using the photometry reduction software \textit{LESVEPHOTOMETRY} V1.2.0.90 \footnote{http://www.dppobservatory.net/} including flat fielding, and bias subtraction.
All of the optical photometric observations are plotted in Figure \ref{fig:AAVOS_lightcurve}.

\subsection{HST/STIS Coronographic Observations}\label{sec:HST_reduction}

HD 163296 was observed with HST/STIS on 2018 April 14 and 2018 July 18 for 4 sequential orbits at each of these epochs. HD 163296 was observed using the STIS occulting bar positions A0.6 (0$\farcs$6 width) and A1.0 (1$\farcs$0 width) for orbits \#1, \#2, and \#4, with each orbit offset in absolute roll angle by 15$^{\circ}$. HST/STIS in coronagraphic mode is filter-less with the CCD sensitive from 2000 to 10,300 \AA.  The (B-V) color-matched PSF star HD 145570 was observed in orbit \#3. The total exposure times for the July epoch were slightly shorter to allow for the entire CCD to be read-out, whereas the April epoch only had the bottom half of the CCD read-out. A summary of these observations are presented in Table \ref{tbl:HST_observations}. We note that while the A0.6 wedge data was taken, reduced, and analyzed, the A0.6 wedge data did not reveal any information that was not seen in the A1.0 wedge, and thus we will not discuss these data any further in this paper.

\begin{table*}
\small
\centering
\caption{HST/STIS Observations}
\begin{tabular}{lcccl}
\hline \hline
{Target Name} & {Date} & {Total} & {Notes} \\
 {} & {} & {Exposure (sec)} & {} \\
\hline
HD 163296 & 1998 Sept. 02 & 864 & Wedge A1.0 \\
HD 141653 & 1999 June 28 & 782 & Wedge A1.0; PSF Star \\
HD 163296 & 2018 April 14 & 4050 & Wedge A1.0 \\
HD 145570 & 2018 April 14 & 896 & Wedge A1.0 ; PSF Star \\
HD 163296 & 2018 July 18 & 3780 & Wedge A1.0 \\
HD 145570 & 2018 July 18 & 896  & Wedge A1.0 ; PSF Star \\
HD 163296 & 2017 July 23 & 1792 & 52'x0.05' slit; PA = 85$^\circ$ \\
HD 163296 & 2017 Aug. 10 & 2021 & 52'x0.05' slit; PA = 45.3$^\circ$ \\
\hline
\end{tabular}
\tablecomments{HST/STIS coronagraphic and UV spectroscopic observing log. The total exposure is the sum total of individual exposures at each epoch. \label{tbl:HST_observations}}
\end{table*}

\begin{figure*}
\centering
\includegraphics[width=\textwidth]{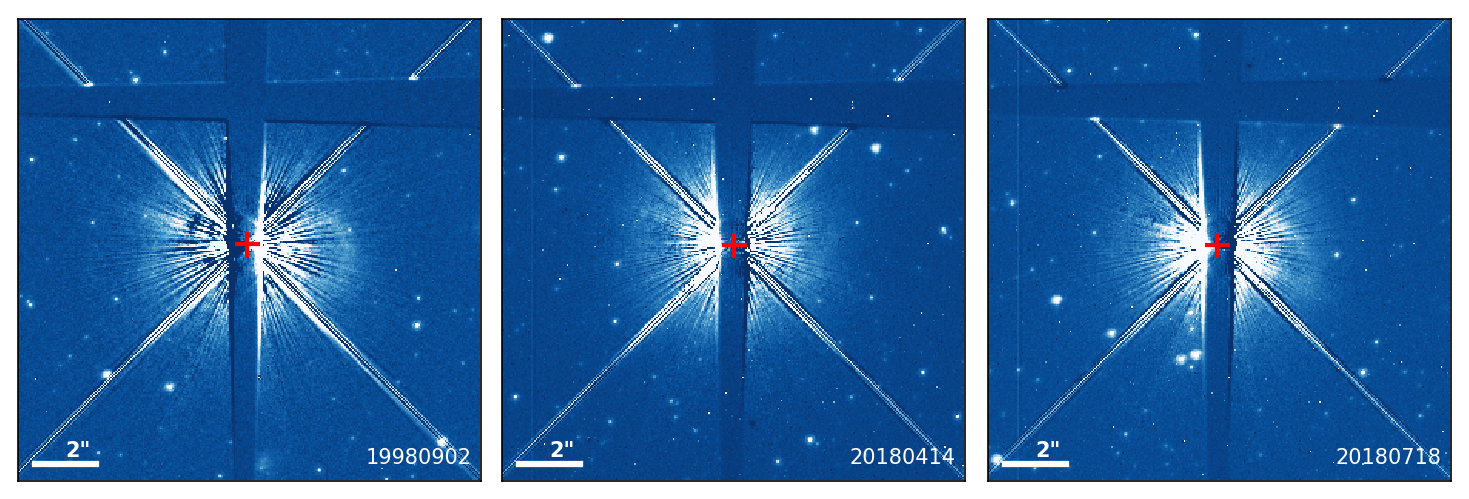}
\caption{PSF subtracted imagery of the HD 163296 disk from 1998 Sept 02 (left), 2018 April 14 (middle), and 2018 July 18 (right) with the A1.0 wedge, spanning a field of view of 15$\farcs$2 $\times$ 15$\farcs$2. The images are all in the detector frame and are unmasked showing the wedges and the spider arms. A red '+' sign marks the location of the central star utilizing the 'x' marks the spot methodology. \label{fig:reduction_examples_large}}
\end{figure*}

We followed the basic PSF subtraction techniques outlined \citet{grady2000}. First, we found the location of the star in each frame using the "x" marks the spot method developed by \citet{schneider2009}. Next, we performed PSF matching, where we subtracted the individual PSF frames from the science images, in turn, to see which PSF frame leaves the smallest residual. This PSF matching technique can help compensate for the "breathing" induced PSF changes HST experiences during a given orbit. Once we found the best PSF match for each science frame, we subtracted the PSF star from the science star, scaling the PSF star flux to compensate for the different apparent brightness. Examples of the PSF subtracted frames are shown in Figure \ref{fig:reduction_examples_large}, where a red cross denotes the location of the star HD 163296. 

We explored how to scale the flux of the PSF star to optimally subtract the central star flux of HD 163296. The nominal V-band magnitude for HD 163296 and the PSF star HD 141653 is V = 6.93 $\pm$ 0.14 mag \citep{ellerbroek2014} and 4.928 $\pm$ 0.009 mag \citep{hog2000} respectively, suggesting a nominal PSF scale factor of 0.158. However, HD 163296 is a known aperiodic variable star (\citealt{ellerbroek2014}; see also Figure \ref{fig:AAVOS_lightcurve}). Thus, we used our AAVSO photometry to help constrain the PSF scaling.
The 2018 Jul epoch observations have contemporaneous AAVSO photometry taken on 2018 July 15 with measured V-band magnitudes of 6.797 $\pm$ 0.013 mag and 6.729 $\pm$ 0.013 (scale factors: 0.179 and 0.190), and the closest photometry points for the 2018 Apr epoch had values of 6.714 $\pm$ 0.011 mag and 6.812 $\pm$ 0.01 mag taken on 2018 April 15 (scale factors: 0.193 and 0.176). Since the STIS observations are clear filter observations, the scale factors are not necessarily perfect, so we explored the scale factor space by performing the PSF subtraction and looking where the most stellar flux has been removed while minimizing the over-subtraction (e.g., negative flux) of the disk. For the Jul epoch, we found the best scaling factor was 0.1773, corresponding to a V-band magnitude for HD 163296 of 6.806 mag. As for the Apr epoch, we found the best scaling factor is 0.1680, corresponding to a V-band magnitude for HD 163296 of 6.87 mag.
These best PSF scaling factors are within 0.1 mag of the contemporaneously observed AAVSO photometry. 

\begin{figure*}
\centering
\includegraphics[width=\textwidth]{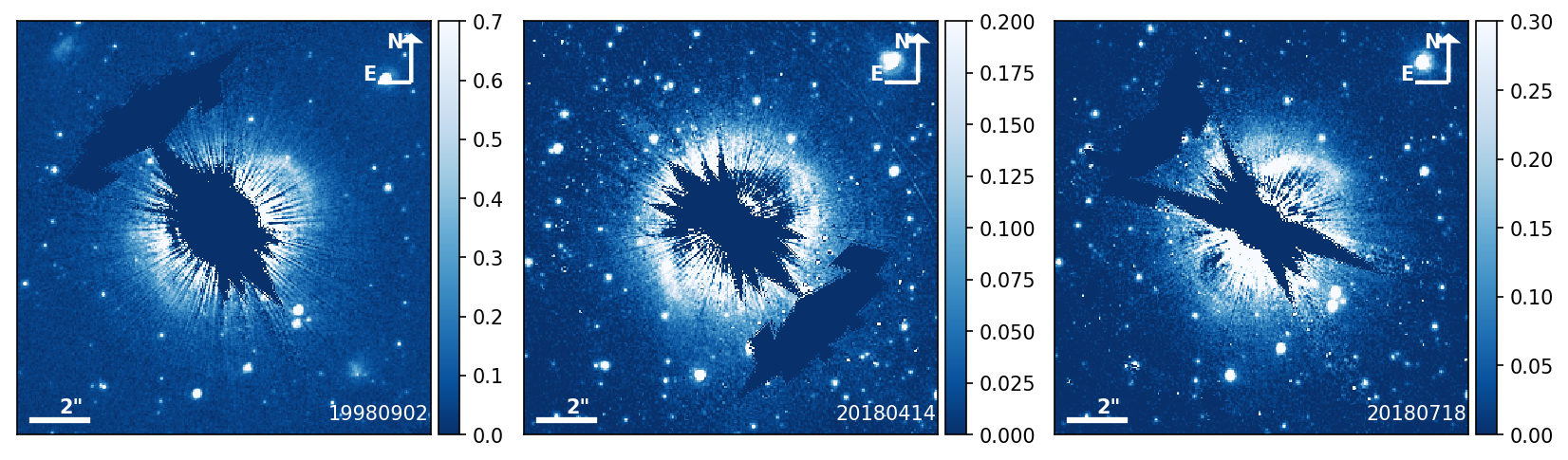}
\caption{Roll-combined, psf-subtracted images of HD 163296 from from 1998 Sept 02 (left), 2018 April 14 (middle), and 2018 July 18 (right). The locations of the coronagraphic wedge and central star have been masked. The images are plotted linearly in units of counts per second, and have a field of view of 15$\farcs$2 $\times$ 15$\farcs$2. \label{fig:images_large}}
\end{figure*}

Finally, the wedge and spider arms were masked, the frames were aligned to a common orientation and rotated north, and median combined to create the final images as shown in Figure \ref{fig:images_large}.
We looked for variation in flux from HD 163296 within a given epoch (2018 Apr or 2018 Jul) and did not detect any significant brightness changes in the disk. 
Note that while the 2018 Apr and 2018 Jul epochs were taken as part of the same observing program, we choose not to stitch these observations together as is traditional (see \citealt{schneider2009}), to enable us to explore potential variability between these epochs, as discussed below in sub-sections \ref{sec:HST_analysis} and \ref{sec:disk_variability}.

We also re-reduced the 1998 Sep epoch observation of HD 163296, which was originally presented in \citet{grady2000}. As noted by \citet{grady2000}, the 1998 Sep epoch did not have contemporaneous observations of a PSF star. We tested several other PSF stars previously observed with HST/STIS and the A1.0 wedge, having similar (B-V) colors as HD 163296 (HD 135298, HD 36546, HD 145570, and HD 141653), to try and reduce the amount of PSF subtraction artifacts in these data. We concluded that the PSF star HD 141653 used in \citet{grady2000} was still the best PSF match. While we did use the same PSF star and the same scale factor as \citet{grady2000}, we adopted a different centering routine ('x' marks the spot), and we used PSF matching when subtracting the best matching individual PSF frame from each science image as described above.
We found the same scaling factor of 0.213 utilized by \citet{grady2000} (PSF V = 5.194 mag, Science V = 6.873 mag) best removed the stellar light from the science images. We visually compared the new final image using our new reduction to that of \citet{grady2000}. We found that we reduced the number of residual speckles around the ansae region of the disk, or the "wagon wheel spoke" effect. We will utilize our new reduction of the 1998 Sep epoch data for the rest of this paper shown in Figure \ref{fig:images_large}.

\subsection{Near IR Data and Reduction}

\begin{figure*}
\centering
\includegraphics[width=\columnwidth]{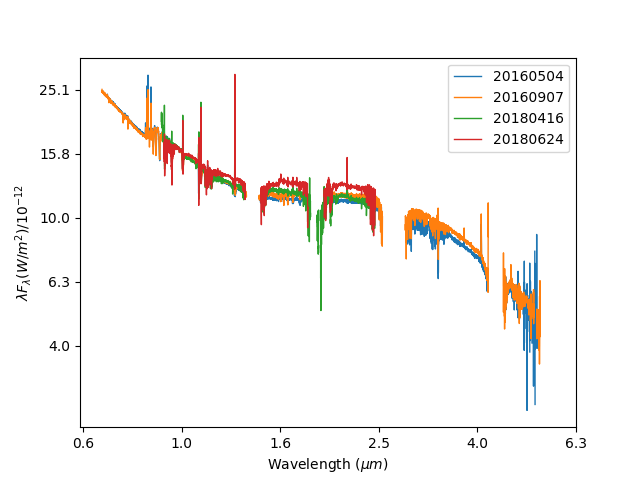}
\caption{Four epochs of flux calibrated near-IR spectra taken between 2016 and 2018 are shown.  Both the highest (2018 June 24) and lowest (2016 May 5) near-IR flux we observed are plotted.  \label{fig:nearIR_spectra}}
\end{figure*}

We monitored the near-IR behavior of HD 163296 from 2016-2018 using NASA's Infrared Telescope Facility (IRTF) and the Apache Point Observatory (APO) 3.5m. We obtained 11 observations from 2016 April to 2018 September using the SpeX spectrograph \citep{rayner2003} at IRTF in its short- (0.8 - 2.4 $\mu m$) and long-wavelength mode (2.3-5.5 $\mu m$) (see Table \ref{tbl:observations}) using a 0$\farcs$8 wide slit. Additionally, we performed 3 observations in 2018 April and 2018 May with the TripleSpec spectrograph \citep{wilson2004} at the APO 3.5m telescope, covering a spectral range of (0.95 - 2.46 $\mu m$) (see Table \ref{tbl:observations}). We observed the A0V star HD 163336 for telluric corrections and flux calibration for all observations. Observations noted in Table \ref{tbl:observations} had contemporaneous IRTF/SpeX prism spectra taken with a 3$\farcs$0 wide slit, which was utilized to correct for absolute flux variations. Observations without prism spectra were scaled to match the observations with prism data based on the optical flux component of their spectra (0.8-0.9 $\mu m$).
These observations were reduced and calibrated using the standard reduction packages \textit{Spextool} and \textit{Triplespectool} respectively \citep{vacca2003,cushing2004}. Note that the near-IR observations on 2018 May 16 and 2018 June 24 were previously presented in \citet{rich2019}.
Sample SpeX and TripleSpec spectra are plotted in Figure \ref{fig:nearIR_spectra}.

\begin{table*}
\small
\centering
\caption{Near-IR and Radio observations.}
\begin{tabular}{lccc}
\hline
\hline \hline
{Telescope/Instrument} & {Target Name} & {Date} & {Spectral Coverage} \\
\hline
IRTF/SpeX & HD163296 & 2016 April 04 & 0.7 - 5.3 ($\mu m$) \\
IRTF/SpeX & HD163296 & 2016 May 04 & 0.7 - 5.3 ($\mu m$)  \\
IRTF/SpeX & HD163296 & 2016 June 09 & 0.7 - 5.3 ($\mu m$)  \\
IRTF/SpeX & HD163296 & 2016 Aug. 10 & 0.7 - 5.3 ($\mu m$)  \\
IRTF/SpeX & HD163296 & 2016 Sept. 07 & 0.7 - 5.3 ($\mu m$)  \\
IRTF/SpeX & HD163296 & 2017 May 25 & 0.7 - 5.3 ($\mu m$)  \\
IRTF/SpeX & HD163296 & 2017 July 27 & 0.7 - 5.3 ($\mu m$)  \\
IRTF/SpeX & HD163296 & 2017 Aug. 14 & 0.7 - 5.3 ($\mu m$)  \\
IRTF/SpeX & HD163296 & 2017 Sept. 12 & 0.7 - 5.3 ($\mu m$)  \\
VLA & HD163296 & 2018 March 06 &  3.76-5.76, 6.36-8.36 GHz  \\
APO/Triplespec & HD163296 & 2018 April 08 & 0.95-2.46 ($\mu m$) \\
VLA & HD163296 & 2018 April 14 & 3.76-5.76, 6.36-8.36 GHz \\
APO/Triplespec & HD163296 & 2018 April 16 & 0.95-2.46 ($\mu m$)  \\
IRTF/SpeX & HD163296 & 2018 April 17 & 0.7 - 5.3 ($\mu m$)  \\
VLA & HD163296 & 2018 May 15 & 3.76-5.76, 6.36-8.36 GHz \\
APO/Triplespec & HD163296 & 2018 May 16 & 0.95-2.46 ($\mu m$)  \\
IRTF/SpeX & HD163296 & 2018 Aug. 11 & 0.7 - 5.3 ($\mu m$)  \\
IRTF/SpeX & HD163296 & 2018 Sept. 22 & 0.7 - 5.3 ($\mu m$)  \\
VLA & HD163296 &  2018 Sept. 27 & 3.76-5.76, 6.36-8.36 GHz \\
VLA & HD163296 & 2018 Oct. 20 & 3.76-5.76, 6.36-8.36 GHz \\
VLA & HD163296 & 2018 Nov. 03 & 3.76-5.76, 6.36-8.36 GHz \\
\hline
\end{tabular}
\label{tbl:observations}
\end{table*}

\subsection{VLA Data and Reduction}

HD 163296 was observed with the VLA using the C-band receiver six times between 2018 March to 2018 November, with three observations in A configuration and three observations in D configuration, as described in Table \ref{tbl:observations}.  The two 1 GHz bandpasses were centered at 4.76 and 7.36 GHz.  
 J1820-2528 was observed to calibrate the complex gain, and 3C286 was observed to calibrate the bandpass and absolute flux density scale. Our observations were intended to be scheduled one month apart and to coincide with our HST/STIS observations; however, due to infrastructure work at the VLA, observations were not possible between May and September of 2018. 
We utilized CASA version 5.1.2 \citep{mcmullin2007} to reduce and analyze these data. Fluxes were measured from cleaned images by fitting a gaussian profile to the flux images using the \textit{imfit} tool. We also preformed RFI flagging. Measured fluxes are listed in Table \ref{tbl:VLA_flux}.

\begin{table*}
\small
\centering
\caption{VLA Flux Densities}
\begin{tabular}{lcc}
\hline
{Epoch} & {Flux Density (Jy)} & {Spectral Index ($\alpha$)}\\
\hline
2018 March 06 & 2.35 $\times 10^{-4}$ $\pm$ 5.63 $\times 10^{-5}$ & 0.817 $\pm$ 0.092 \\
2018 April 14 & 1.873 $\times 10^{-4}$ $\pm$ 8.60$\times 10^{-5}$ & 0.715 $\pm$ 0.082 \\
2018 May 15 & 2.299$\times 10^{-4}$ $\pm$ 7.15$\times 10^{-5}$ & 0.708 $\pm$ 0.021 \\
2018 Sept. 27 & 3.434$\times 10^{-4}$ $\pm$ 3.23$\times 10^{-5}$ & 1.0104 $\pm$ 0.0488 \\
2018 Oct. 20 & 3.065$\times 10^{-4}$ $\pm$ 2.61$\times 10^{-5}$ & 1.080 $\pm$ 0.0246 \\
2018 Nov. 03 & 3.164$\times 10^{-4}$ $\pm$ 2.83$\times 10^{-5}$ & 1.219 $\pm$ 0.029 \\
\hline
\end{tabular}
\label{tbl:VLA_flux}
\end{table*}

\subsection{UV Data and Reduction}

\begin{figure*}
\centering
\includegraphics[width=\textwidth]{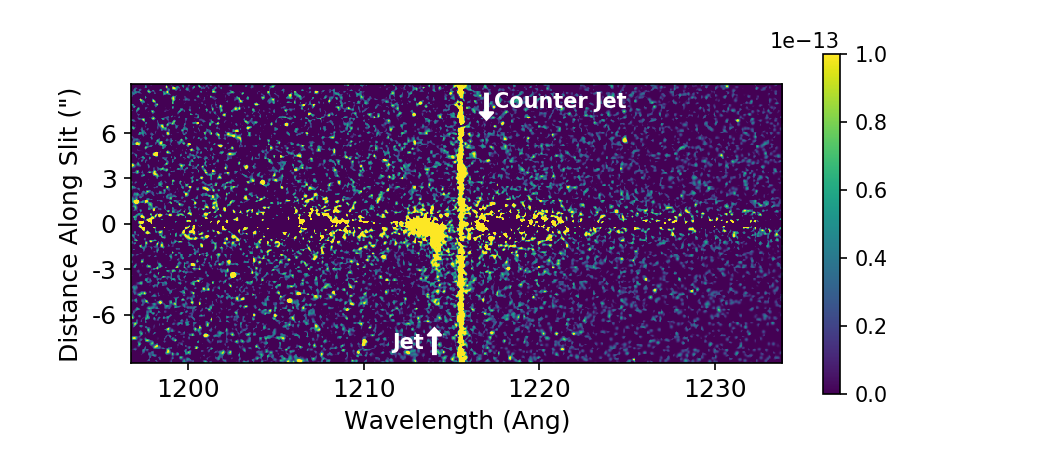}
\caption{A HST/STIS UV observation of HD 163296 centered around the lyman alpha line (1215 \AA).  Here the orientation of the long-slit (45.3$^\circ$) was nearly aligned with the jet axis (42.5$^\circ$), and the background and stellar contributions were minimized by subtracting an observation with the long-slit at a PA of 85$^\circ$. The predicted locations of the jet (towards us) and counter jet (away from us) are labeled with arrows. While we can see marginal evidence of the jet, we see no evidence for the counter jet, the latter of which could be due to the long-slit not being perfectly aligned with the PA of the jet axis. \label{fig:jet_UV}}
\end{figure*}

HD 163296 was observed with HST/STIS using the FUV-MAMA detector, the G140M grating centered 
at 1218 \AA, and the 52'x0.05' slit as part of program 14690 (G\"{u}enther PI). Two observations occurred on 2017 July 23 and 2017 August 10 at slit position angles of 85$^\circ$ (approximately perpendicular to the jet axis) and 45.3$^\circ$ (approximately aligned with the jet axis of 42.5$^\circ$), and exposure times of 1792 sec. and 2021 sec. respectively. Both observations were reduced using the standard STIS pipeline, with standard reduction techniques that included dark-subtraction, flat-fielding, and wavelength calibration. Subtracting the 2017 July epoch from the 2017 Aug. epoch, scaled to match exposure depth, allows us to remove background flux and star continuum flux without affecting the jet. The resultant 2D spectra is shown in Figure \ref{fig:jet_UV}. 

\section{HST/STIS Analysis}\label{sec:analysis}
\subsection{Disk geometry}\label{sec:HST_analysis}
Scattered light from HD 163296's disk is observed in all of the HST/STIS epochs, as shown in Figure \ref{fig:images_large}, from the edge of our effective inner working angle $\sim$0$\farcs$5 (51 au) to 5$\farcs$ (500 au). The disk major axis is oriented approximately at PA of 132$^\circ$ similar to what has previously been observed with ALMA \citep{isella2016} and is inclined at 42$^\circ$. In the re-reduced 1998 Sept epoch data, we observe the two ansae (SE and NW ansae) originally presented by \citet{grady2000}, located at a projected distance of 3$\farcs$3 (330 au) from the star. The disk ring flux distribution is brightest on the near side of the disk (NE-side) and we do not see any flux on the far side of the disk (SW-side). This flux distribution is similar to the broken ring structures seen at the 1st ring (0$\farcs$65, 66 au) in near-IR observations \citep{garufi2017,monnier2017,muro2018,rich2019}. While we refer to these features as ansae, we are interpreting them as the 4th dust ring in the HD 163296 system. We note that both ansae were not observed simultaneously in our 2018 dataset. We only observed the SE ansae in the 2018 April epoch data, and we only observed the NW ansae in the 2018 July epoch data. Previously, \citet{wisniewski2008} had also only observed one ansae being present around HD 163296. The location of the ansae had not changed from the 1998 epoch and the 2018 epoch.

\begin{figure*}
\centering
\includegraphics[width=0.5\textwidth]{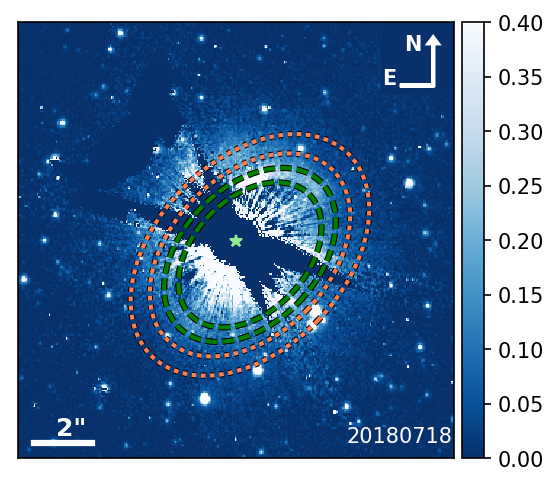}
\caption{A 15$\farcs$2 $\times$ 15$\farcs$2 view of the 2018 July HST imagery.  Surface brightness profiles along the ansae region (green ellipses located at 300 and 360 au) are presented in Figure \ref{fig:disk_ansae_large}, while surface brightness profiles along the disk periphery region (orange ellipses located at 420 and 500 au) are presented in Figure \ref{fig:disk_excess_large}. \label{fig:disk_example_ellipse}}
\end{figure*}

Assuming that the broken ring is circular, has an inclination of 42$^\circ$, and PA of 132$^\circ$ \citep{isella2016}, we plotted an ellipse on the 1998 Sept epoch image (Figure \ref{fig:disk_example_ellipse}) and found that the broken ring disk is consistent with a minor axis offset of 0$\farcs$7. The 2018 April and 2018 July observations are also consistent with the same minor axis offset. Assuming that the 4th ring is perfectly centered around the star, the minor axis offset of 0$\farcs$7 would correspond to a dust scale height of 64 au at a radial distance of 330 au. We will discuss the minor axis offset for HD 163296 in Section \ref{sec:discussion}.

\begin{figure*}
\centering
\includegraphics[width=\columnwidth]{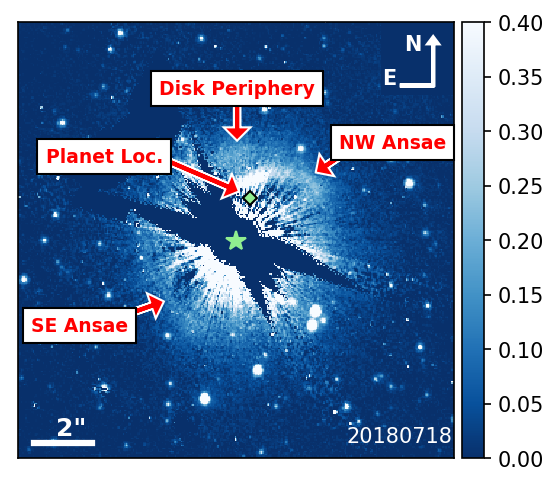}
\caption{A 15$\farcs$2 $\times$ 15$\farcs$2 view of the 2018 July epoch imagery, with features such as wthe NW Ansa, SE Ansa, disk periphery flux, central star location (green star), and the approximate location of \citet{pinte2018} planet candidate (green diamond) annotated. \label{fig:disk_label}}
\end{figure*}

We observe additional non-azimuthal scattered light from the near-side of the disk north of the NW ansa, increasing in intensity as it approaches the NE minor axis. This feature is observed in all three epoch images (Figure \ref{fig:images_large}), and is labeled as ``disk periphery'' in Figure \ref{fig:disk_label}.  It has a larger radial extent than the prominent ansae. We will discuss the origin of this feature later in Section  \ref{sec:discussion}. Finally, the middle sub-image in Figure \ref{fig:images_large} showing epoch 2018 Apr, there is a faint linear streak radiating from the top of the disk at a PA of 45$^\circ$ along the jet axis. While, this could be a signature of the jet, the linear streak resembles other radial streaks observed in the 2018 Apr and Jul epochs. Thus, the feature is most likely a residual.

\subsection{Disk Surface Brightness}\label{sec:disk_assymetry}

\begin{figure*}
\centering
\includegraphics[width=\textwidth]{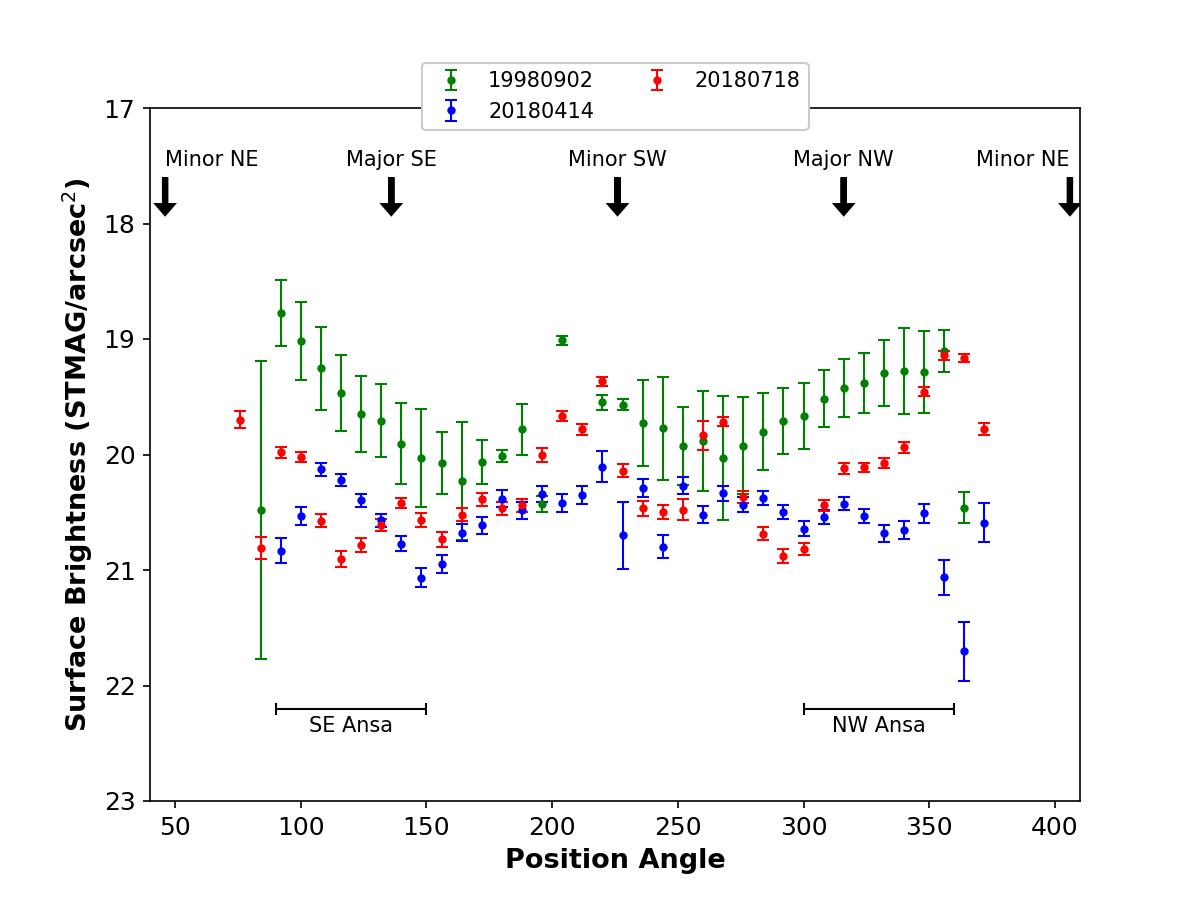}
\caption{The azimuthal surface brightness of the ansae (4th ring) region, binned 8$^\circ$ wide in azimuthal angle and 2$\farcs$96 - 3$\farcs$55 (300 - 360 au) in projected radial space. We assumed that the disk ring is circular and adopted an inclination of 42$^\circ$ and PA of 132$^\circ$ from \citep{isella2016}. The error bars plotted are 3$\sigma$, as discussed in the text. \label{fig:disk_ansae_large}}
\end{figure*}

We computed the azimuthal dependence of the disk surface brightness within the ansae region of the disk (4th ring; 300-360 au; see Figures \ref{fig:disk_example_ellipse} and \ref{fig:disk_label}), binning the flux within these elliptical annuli using an assumed inclination of 42$^\circ$ and PA of 132$^\circ$ \citep{isella2016}. The resultant azimuthal surface brightness profile (Figure \ref{fig:disk_ansae_large}) includes 3-$\sigma$ error bars computed from photon noise, read noise, and dark noise contributions.
Several azimuthal bins are contaminated by background stars causing these regions to have abnormally high flux, such as the region near the SE minor axis.

Figure \ref{fig:disk_ansae_large} exhibits a large variation in the surface brightness of the ansae region of the disk both azimuthally (position angle) and in time. In the 1998 Sept epoch, the disk is brighter closer to the NE minor axis than the SW minor axis. The SE (19.65 $\pm$ 0.11 STMag arcsec$^{-2}$ (PA=124$^\circ$)) and NW ansae (19.29 $\pm$ 0.10 STMag arcsec$^{-2}$ (PA=332$^\circ$)) exhibit similar surface brightnesses. By contrast, during the 2018 April epoch, we only detect the SE ansa, and its surface brightness (20.40$\pm$0.02 STMag arcsec$^{-2}$; PA= 124$^\circ$) is $\sim$ 0.75 STMag arcsec$^{-2}$ dimmer than in 1998 Sept. The surface brightness near the expected location of the NW ansa is flat in 2018 April and much dimmer (20.68$\pm$0.02 STMag arcsec$^{-2}$; PA=332$^\circ$) than the 1998 Sept epoch.  Conversely, we only detect the NW ansa in the 2018 July epoch.  The surface brightness of the NW ansa (20.08$\pm$0.01 STMag arcsec$^{-2}$; PA=332$^\circ$) is $\sim$0.79 STMag arcsec$^{-2}$ dimmer than then 1998 Sept epoch. The surface brightness near the expected location of the SE ansa in the 2018 July epoch is dimmer (20.78$\pm$0.02 STMag arcsec$^{-2}$; PA=124$^\circ$) than both the 1998 Sept and 2018 April epochs. We will discuss the implications of these results in subsection \ref{sec:disk_variability}.

\begin{figure*}
\centering
\includegraphics[width=\textwidth]{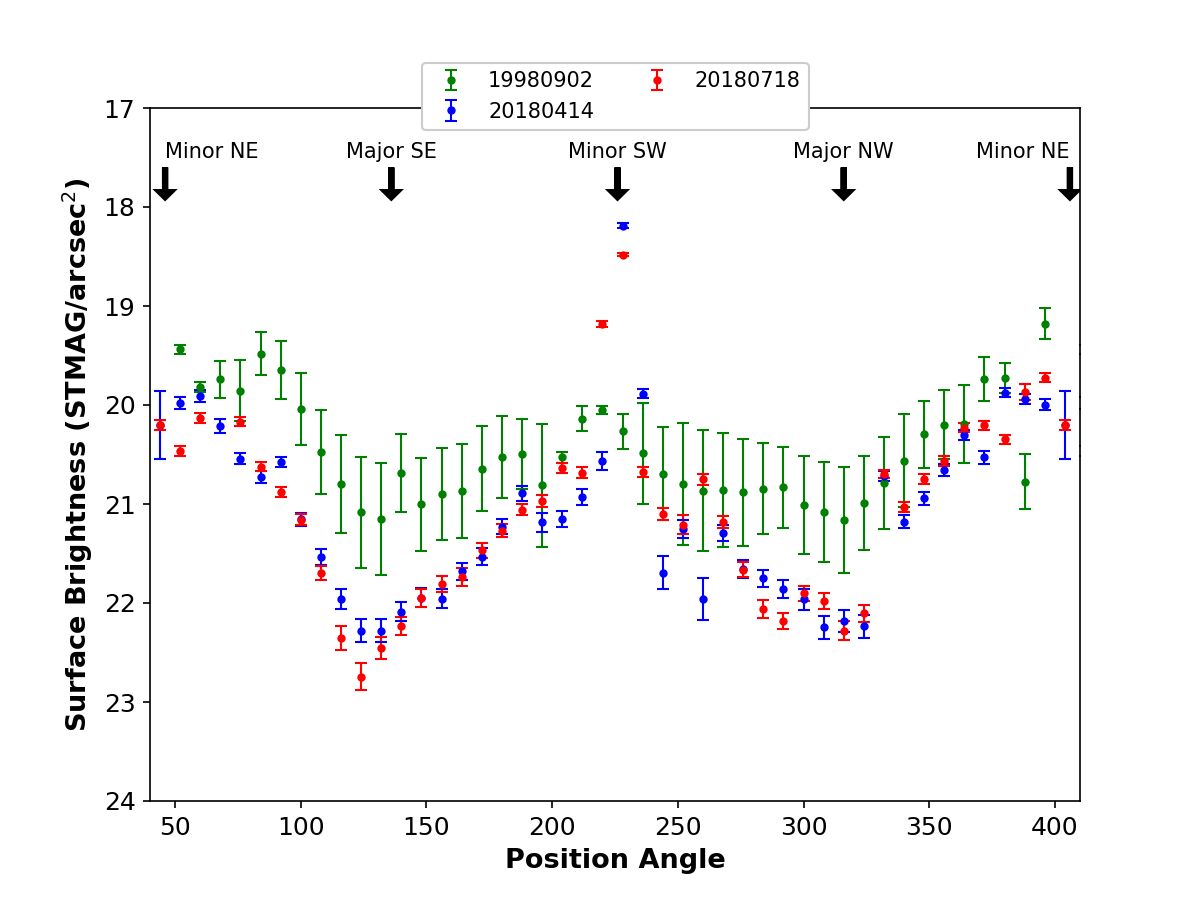}
\caption{The azimuthal surface brightness of the disk periphery region, binned 8$^\circ$ wide in azimuthal angle and 4$\farcs$1 - 4$\farcs$9 (420 - 500 au) in projected radial space. We assumed that the disk ring is circular and adopted an inclination of 42$^\circ$ and PA of 132$^\circ$ from \citep{isella2016}. The error bars plotted are 3$\sigma$, as discussed in the text. \label{fig:disk_excess_large}}
\end{figure*}

We similarly computed the azimuthal surface brightness of the disk periphery region (420-500 au; see Figures \ref{fig:disk_example_ellipse} and \ref{fig:disk_label}), as seen in Figure \ref{fig:disk_excess_large}. We note that all epochs exhibit the same trend, the surface brightness increases towards the NE minor axis (i.e. the near-side of the disk), and we observe little azimuthal variation between the two 2018 epochs. However, the 2018 April (20.54$\pm$0.02 STMag arcsec$^{-2}$) and 2018 July (20.17$\pm$0.02 STMag/$"$) epochs are dimmer than the 1998 Sept epoch (19.85 $\pm$ 0.10 STMag arcsec$^{-2}$) at a PA= 12$^\circ$). Figure \ref{fig:disk_excess_large} exhibits a large jump in surface brightness along the SW minor axis, that is caused by background stars within those bins. An additional $\sim$1 STMag arcsec$^{-2}$ discontinuity in the surface brightness is present in the 2018 epochs at PA= 330$^\circ$. This is further evidence of non-azimuthal variations within the disk and possibly due to a transient shadowing effect that was not present during the 1998 Sept epoch.

\subsection{HH-knots}

\begin{figure*}
\centering
\includegraphics[width=\columnwidth]{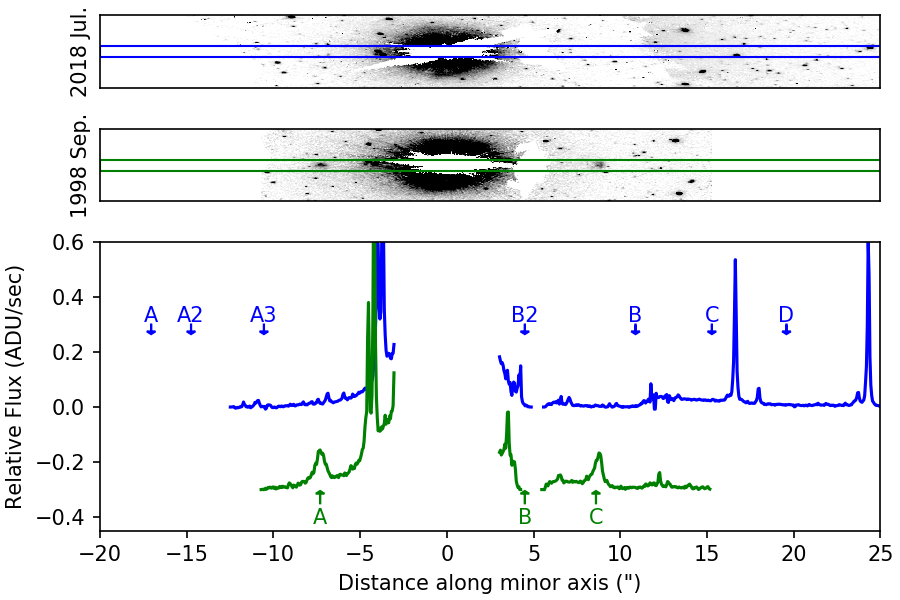}
\caption{Crosscut along the jet axis from the 1998 Sept (green) and 2018 July (blue) HST observations. The HH knot locations are provided for their measured locations in the 1998 Sept epoch and the projected location in the 2018 data from \citet{ellerbroek2014}. The top two panels respectively shows rotated cutout images of the 1998 Sept and 2018 July coronagraphic imaging epochs also shown in Figure \ref{fig:disk_ansae_large}. While we observe clear evidence of two HH-knots (A and C) in the 1998 Sept data (green), we do not see any of the 5 HH-knots predicted to be observed in the 2018 July epoch (blue). The large thin spikes located around HH-knots C and D are background stars and seen in the top panel. \label{fig:HH_knot}}
\end{figure*}

\citet{grady2000} and \citet{devine2000} identified three HH-knots, labeled A, B, and C, along the minor axis of their 1998 Sept HST observations (see Figure  \ref{fig:images_large}). 
To date there have been 11 HH-knots associated with HD 163296 (A,A2,A3,B,B2,C,D,E,F,G, and H; \citealt{grady2000,devine2000,wassell2006,ellerbroek2014}).  Given the measured proper motion of the HH-knots \citep{ellerbroek2014}, (Red Jet: $v_{t,red} = 0.28 \pm 0.01 "/yr$, Blue jet: $v_{t,blue} = 0.49 \pm 0.01 "/yr$), we predict five of these HH-knots (A3,B2,B,C, and D) should fall within the FOV of our 2018 July epoch data. 
Utilizing the HH-knot proper-motion analysis from \citet{ellerbroek2014} and citations therein, we annotate the predicted locations of the HH-knots for the 1998 Sept and 2018 July HST data (Figure \ref{fig:HH_knot}). 
Surprisingly, we do not see any HH-knots in either our 2018 April or 2018 July observations. 
We inserted artificial HH-knots into our data at the predicted locations of HH-knots B and C, performed aperture photometry, and found that the knots could be 15 times dimmer and be detected with a 3-$\sigma$ significance above the background. We will discuss the non-detection of these HH-knots in sub-subsection \ref{sec:jet}.

\subsection{Jet in UV}

Figure \ref{fig:jet_UV} shows the clear presence of the jet associated with HD 163296 in the 2017 UV observation, out to a distance of 5$\farcs$0, but do not reveal any evidence of the counter jet. Since the position angle of the long slit was not perfectly aligned with the jet axis and the slit is four times smaller (0$\farcs$05) than previous HST/STIS long-slit observations of the jet (0$\farcs$2: \citealt{devine2000}), we expect much of the jet flux at larger distances from the star to be located outside of the slit. For this reason, we do not compare the flux of the jet that we observe to previous studies \citep{devine2000, wassell2006, gunther2013}. We expect the counter jet to be masked by the forward portion of the protoplanetary disk which extends out along the minor axis of the disk to a projected distance of 2$\farcs$9. Thus a portion of the counter jet could have been detected. We measure that the jet has a velocity of -336 km s$^{-1}$, which matches what was previously observed values by \citet{devine2000} (380 - 330 km s$^{-1}$).

\section{Analysis of Supporting Multi-wavelength Observations}\label{sec:analysis_mult}

We performed a multi-wavelength monitoring campaign in the optical, near-IR, and radio wavelengths to try and detect the predicted launch of an HH-knot from HD 163296. Below we describe the optical, near-IR, and radio variability of these data.

\subsection{Optical Variability}

Our optical (B-, V-, and I-band) photometry of HD 163296 throughout 2018 (Figure \ref{fig:AAVOS_lightcurve}) exhibits evidence of moderate variability. After excluding discrepant photometry using a 3-$\sigma$ clipping algorithm, we find the average magnitude of the system to be 6.85 $\pm$ 0.15 mag, 6.75 $\pm$ 0.11 mag, and 7.27 $\pm$ 0.09 mag in the B-, V-, and I- filters respectively.
The typical error for an individual observation is 0.02 mag, 0.01 mag, and 0.02 mag respectively in these filters. 
The clipping algorithm is necessary as there are discrepant photometry data points that could induced by non-photometric skies. Without observing logs, we cannot strategically remove these points any other way. We note that this analysis could be removing high-amplitude variations in flux in the light curve.

We observe two potential dipper events on 2018 June 7 and 2018 August 7, that each appear in all three band-passes. The first event, which occurred between the two epochs of our HST coronagraphic observations, is immediately proceeded by a gradual rise to the average brightness over a several day timeframe. The event had an amplitude of $\sim$0.5 magnitudes, although this is only a lower limit due to the sparse temporal sampling of our data. The 2018 June 7 event appears similar to 4 previous events noted by \citet{ellerbroek2014}, and is both smaller and shorter in duration than the 2001 event that reduced the system flux by $\sim$ 1 magnitude and lasted several months \citep{ellerbroek2014}. The second dipper event was only observed by a single night observation, albeit in three filters. Robustly quantifying both of these dipper events is hampered by the sparse temporal coverage of our data. Thus we label the first event as potential dipper event, and the second as a suggestive dipper event. 
We note that further analysis of the color of dipper events in 2012 of this system is presented in the appendix of \citet{pikhartova2019}.

\subsection{Near-IR Variability}

\begin{figure*}
\centering
\includegraphics[width=\columnwidth]{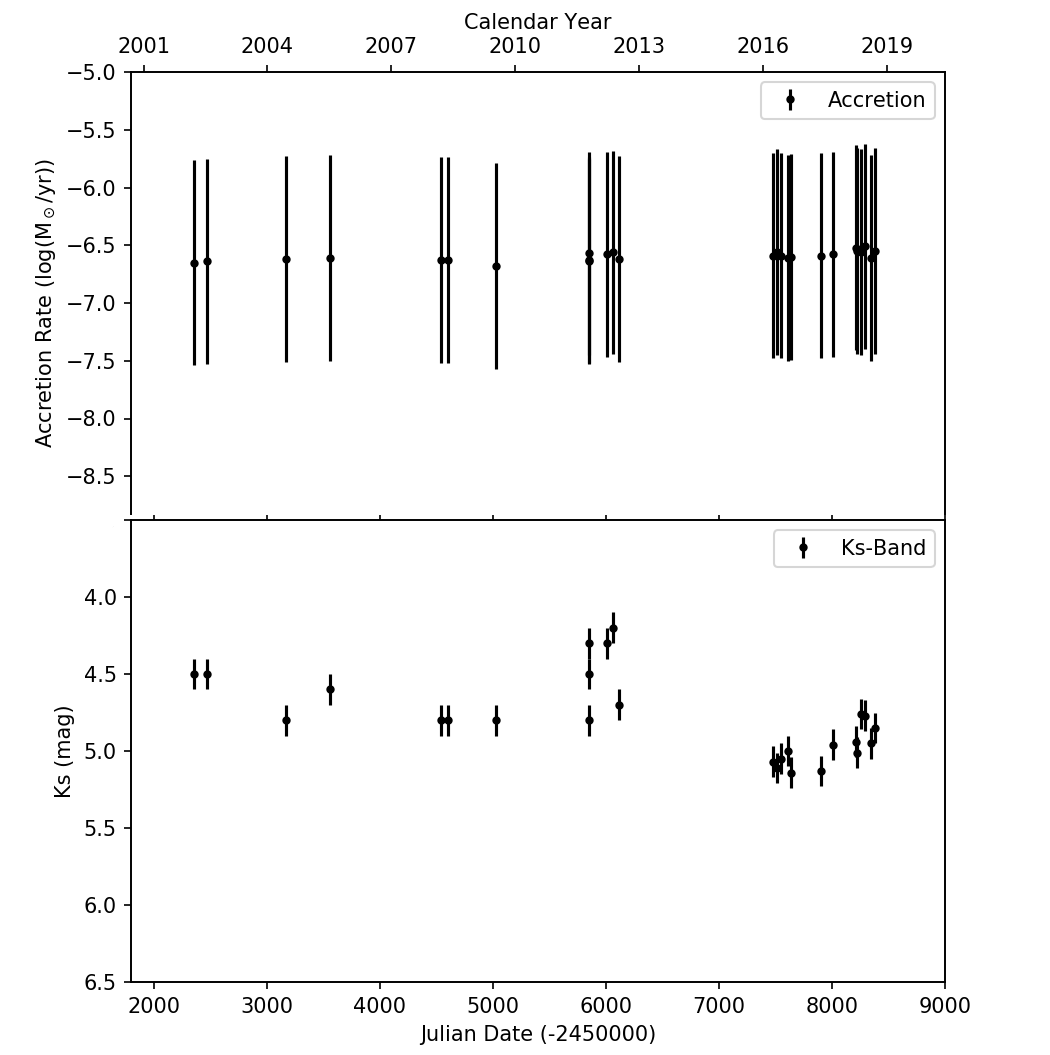}
\caption{The top panel shows the accretion rate for HD 163296 measured with the Br$\gamma$ line. Accretion rates before 2013 were originally presented by \citet{ellerbroek2014} and were recalculated by this work to incorporate new distance and $\Rsun$ values. The bottom panel shows the lightcurve in the Ks-band. Ks-band magnitudes before 2013 were taken from \citet{ellerbroek2014} (see Table \ref{tbl:accretion}). \label{fig:Accretion_lightcurve}}
\end{figure*}

We convolved a K-band filter with our near-IR flux calibrated spectra, to extract K-band magnitudes and quantify variability in these data (see Table \ref{tbl:accretion} and Figure \ref{fig:Accretion_lightcurve}). We have also included historical K-band magnitudes from \citet{ellerbroek2014} and citations therein in Figure \ref{fig:Accretion_lightcurve}. 
The K-band light curve exhibits variations on a timescales of both months and years, with a total amplitude of $\sim$0.94 magnitudes. Although our temporal coverage between 2012 and 2016 was poor, the system clearly dimmed by $\sim$0.5 magnitudes during this time-frame.  Since 2016, the system has exhibited a gradual $\sim$0.4 magnitude increase in Ks-band flux.  We do not see any outburst events in these data, such as the 2002 outburst that possibly coincided with the launch of a HH-knot \citep{ellerbroek2014,pikhartova2019}.

\subsection{Accretion History of HD 163296}

We calculate the mass accretion rate outlined in \citet{ellerbroek2014}. Equivalent widths (EW) of the 
Br$\gamma$ line were measured using \textit{IRAF} \textit{splot} tool using a gaussian function (Table \ref{tbl:accretion}). 
To calculate the line luminosity, we first correct our measured EWs$_{\textrm{obs}}$ by removing the effect of the photospheric line, as shown in Equation \ref{eqn:EW_correct}. 
We adopt the same photospheric EW (EW$_{\textrm{phot}}$ = -22 \AA) as \citet{mendigutia2013} and the same reddening (A$_V$ = 0.5 mag) as \citet{ellerbroek2014}. 
We use these values to calculate the line luminosity (L$_{\textrm{line}}$; Equation \ref{eqn:line_lum}). We then converted the line luminosity into the accretion luminosity in Equation \ref{eqn:lum_acc} utilizing the relation by \citet{fairlamb2017}. 
Finally, the mass accretion rate is calculated in Equation \ref{eqn:mass_acc} assuming an R$_{\ast}$ = 1.7 $\Rsun$  ~ \citep{pikhartova2019}. The values are listed in Table \ref{tbl:accretion} and plotted in Figure \ref{fig:Accretion_lightcurve}. 
We recalculated the accretion rates previously presented in \citet{ellerbroek2014} and included the updated distance and R$_{\ast}$ values, and included an updated line luminosity into the accretion luminosity relation. 
We find no substantial changes in accretion rate from 2003 to 2018, as shown in Figure \ref{fig:Accretion_lightcurve}.\\

\begin{equation}\label{eqn:EW_correct}
    EW_{\textrm{cs}} = EW_{\textrm{obs}} - EW_{\textrm{phot}} 10^{-0.4 \mid \Delta K \mid}
\end{equation}

\begin{equation}\label{eqn:line_lum}
    L_{\textrm{line}} = 4\pi d^2 EW_{cs} F_K 10^{0.4A_K}
\end{equation}

\begin{equation}\label{eqn:lum_acc}
    \textrm{log}(\frac{L_{\textrm{acc}}}{L_\odot}) = 4.46(\pm 0.23) + 1.30(\pm 0.09) \times \textrm{log}(\frac{L_{\textrm{Br} \gamma}}{L_\odot})
\end{equation}

\begin{equation}\label{eqn:mass_acc}
    \dot{M}_{\textrm{acc}}^{\ast} = \frac{L_{\textrm{acc}}R_{\ast}}{G M_{\ast}}
\end{equation}

\subsection{Radio Flux Variability}

Fluxes from the C-band VLA observations are listed in Table \ref{tbl:VLA_flux}.  During the launch of a HH-knot, we expect to observe the radio flux increase by 50\%, based on previous measurements of HH-knot launches \citep{devine2000}. 
Additionally, we should observe a change in the spectral index, with the value possibly becoming negative. We do not observe any significant changes in the C-band flux to suggest the launch of an HH-knot. Additionally, our spectral index values are positive, which is consistent with flux originating as thermal free-free emission. Our radio observations are consistent with no new HH-knots being launched during 2018. 

\section{Discussion}\label{sec:discussion}
\subsection{Disk Structure and Illumination}\label{sec:disk_variability}

The outer disk, as imaged with ALMA, is a series of three concentric rings at $0 \farcs 66$, $1 \farcs 0$, and $1 \farcs 6$ (67 au, 102 au, and 160 au; \citealt{isella2018,dent2019}) with gaps in between each ring. We observed two ansae that form a 4th ring at 3$\farcs$25 (330 au). Assuming this ring is circular, we measured a minor axis offset value of 0$\farcs$7, which would correspond to a dust scale height of 64 au (0$\farcs$7 offset). This latest dust scale height measurement joins a host of other minor axis offset measurements, such as those of the first dust ring in the near-IR (0$\farcs$0432 $\pm$ 0$\farcs$0016, \citealt{rich2019}; 0$\farcs$06, \citealt{garufi2014}; 0$\farcs$105 $\pm$ 0$\farcs$045, \citealt{muro2018}; 0$\farcs$1, \citealt{monnier2017}). As we add more and more measurements, we can better map the vertical dust distribution as a function of stellocentric separation, which will aid modeling of the system. 

It has been previously noted that HD 163296 exhibits azimuthally asymmetric disk illumination on timescales $<$ 4 years \citep{wisniewski2008,rich2019}. Our observations (Section \ref{sec:disk_assymetry}) indicate the timescale for such illumination variations are $<$ 3 months. Large-scale, azimuthally asymmetric illumination has been observed both for the innermost disk ring ($0 \farcs 66$, 67 au \citealt{rich2019}) and now the fourth disk ring (3$\farcs$25, 330 au; \citealt{wisniewski2008} and current work), suggesting that a similar mechanism is responsible for these phenomenon. As discussed in \citet{rich2019}, such mechanisms could include a warped inner disk structure shadowing the outer disk \citep{sitko2008}, or dust ejected above the mid-plane of the disk that shadows the outer disk \citep{ellerbroek2014}.

Assuming that the material causing the illumination variations is in a keplerian orbit around the star, the observed $<$ 3 month variability timescale and a stellar mass of 2.14 $\Msun$ \citep{pikhartova2019} implies a semi-major axis $<$ 0.5 au. This is consistent with the nominal outer edge of the inner disk, $~$0.41 au \citep{setterholm2018}.  If the source of the shadowing resides in the inner disk, this material must reach a scale height of at least 0.08 au above the mid-plane in the inner disk (at 0.41 au) to shadow the 4th dust ring. 
MCRT modeling by \citet{pikhartova2019} has suggested that dust located interior to 0.5 au is ejected above the disk mid-plane to a scale height of at least 0.08 au, and this could account for the broad-band optical and near-IR photometric variability observed in the system's 2002 outburst. If this ejected dust is not azimuthally symmetric, it could also explain the azimuthally asymmetric disk shadowing that has been observed in the first and 4th dust rings.

We can also estimate the azimuthal size of source of the disk shadowing. Based on Figure \ref{fig:disk_ansae_large}, we estimate that the ansae in the 4th ring are at least $\sim$30$^\circ$ in azimuthal extent. Assuming the source material that creates the shadowing resides $<$ 0.5 au from the star, this material must subtend an azimuthal size of 0.26 au to create a $\sim$30$^\circ$ shadow in the 4th ring.

\subsection{Optical and Near-IR light curve variability}

The optical and near-IR light curves, Figures \ref{fig:AAVOS_lightcurve} and \ref{fig:Accretion_lightcurve} respectively, do not exhibit any large dips or bursts indicative of the launch of any new HH-knots within our observing windows.
The optical light curve exhibits two dipper events: one likely event on 2018 June 7 and one possible event on 2018 August 7. The 2018 June 7 event has a depth of at least 0.52 mag in V-band. The light curves also exhibit variations larger than individual photometric uncertainties, suggesting that additional short timescale variations could be present.

Dippers have been attributed to YSO's with inclinations that are nearly edge on, with fluctuations of the inner disk scale height causing dimming of the star \cite{cody2014}.  More recently, it has been shown that dipper events can occur in systems where the outer disk has at a low to moderate inclination (0$^\circ$ - $\sim$60$^\circ$), and even might be occurring independently of inclination \citep{Ansdell2019}. 
HD 163296 is one such system that has a moderate outer disk inclination (42$^\circ$) and experiences dipper events (see Figure \ref{fig:AAVOS_lightcurve} and \citealt{ellerbroek2014}). The disk's inclination of 42$^\circ$ suggests that for the star to be occulted by disk material, dust would have to reach a height of 0.37 au above the disk at the nominal inner disk radius (0.41 au). 

We can estimate the azimuthal size of the feature responsible for the dipper, using its observed duration. Based on the 2018 June 7 event, the duration of the event cannot be longer than 10 days. This duration is similar to some of the dipper events observed in the system by \citet{ellerbroek2014}.  If we assume that the dust causing the event is located around the inner disk (0.41 au) and is moving at Keplerian speeds, we find that the diameter of the occulting material is 0.3 au. This corresponds to an azimuthal angle of 27$^\circ$. 

We remark that the physical sizes we have derived for inner disk ($<$ 0.5 au) features that could cause both dipper events and large azimuthally asymmetric shadowing of the first and fourth disk rings are broadly consistent with one another.  These similarities suggest that the same (or similar) mechanisms could be responsible for both types of observational phenomenon in the system.

Finally, we note there is possible long term variability in our near-IR flux observations. Our observations at the beginning of 2016 are $\sim$0.5 mag dimmer and the K-band flux increases by 0.3 mag from 2016-2018. 
Long term variability in the IR light curves of disk hosting stars has previously been found by \citet{cody2014}. We do not observe similar long term variations in the optical lightcurve. Multi-wavelength modeling of these lightcurves are required to assess the origin of the different observed behaviors.

\subsection{Jet Activity}\label{sec:jet}

We have not detected the launch of a new HH-knot as of 2018 Nov 3. \citet{ellerbroek2014} predicted that the system should launch a knot every 16 $\pm$ 0.7 yr, calculated from the proper motion and radial velocity of current HH-knots. Since the last known launch of an HH-knot was $\sim$2002, a new HH-knot was expected in $\sim$2018. \citet{ellerbroek2014} also predicted that there should be a decrease in optical flux from the reddening due to the disk wind and an increase in the near-IR flux from thermal radiation from dust in the disk wind. 
Instead, we see two minor optical dimming events in 2018 (see Figure: \ref{fig:AAVOS_lightcurve}) and a general decrease in near-IR flux from 2016-2018 as compared to Ks-band photometry before 2013. 

We can not definitively rule out the 16 year periodicity of knot launches, as there is a $\sim$32\% statistical probability that such an event could have happened after our final observation in Nov 2018. Imaging and/or spectroscopic observations beyond the 3-sigma uncertainty of the periodic behavior suggested by \citet{ellerbroek2014} (e.g. in 2020 or later) that are capable of identifying new HH-knots are recommended to confirm our non-detection.

We do not detect any of the previously launched HH-knots in our 2018 April and 2018 July HST/STIS imaging.
\citet{gunther2013} notes if a HH-knot is not shock-heated, the knot itself can cool on a timescale of $\tau$ = 0.4 yr. Thus it is possible that the HH-knots have cooled sufficiently such that they are no longer visible. We note that the lack of detection of HH -knots does not necessarily correlate with a change in the activity of the jet itself. 
The jet associated with HD 163296 has a decades long history of activity. As the HH-knots are primarily fed energy through shock heating and cool on relatively short timescales, the HH-knots could be interacting with less material than previous HH-knots and they have cooled sufficiently to no longer be detectable with our 2018 Apr and Jul epoch imagery.
Future observations to directly measure the current jet activity and confirm the dimming of older HH-knots, e.g. with deeper ground-based slit-spectroscopy, would help test our interpretation of current data.

The last detection of the jet in Lyman $\alpha$ was 2017 Aug 10. We do not see any emission beyond 5$\farcs$0; however, this is most likely due to the jet being located outside of the slit. We also do not see any evidence of any of the HH-knots in these data. The time-lag between this UV jet detection and the 2018 epoch coronagraphic observations are far enough apart for significant cooling to have occurred. 

\subsection{Effects of Planet Candidate on Disk}\label{sec:planet}

\citet{pinte2018} inferred the presence of a 2 M$_{jup}$ mass planet located at 260 au based on analysis of gas velocities. We do not observe any excess flux in this region of our image, which is expected as such a planet is too dim to be detected with our HST/STIS observations. We can look for the effects of the planet on the disk. In Figure \ref{fig:disk_label}, we have placed a diamond at the approximate separation and position angle of the proposed planet. Interestingly, the planet is located between the 3rd ring (160 au) and the ansae region (4th ring; 330 au). Much like the 1st and 2nd rings \citep{teague2018}, the 4th ring could be formed through the dynamics of the \citet{pinte2018} planet. New dynamic modeling of the disk that includes the \citet{pinte2018} planet and the two proposed planets by \citet{teague2018} in the two inner gaps, and replicates the shadowing phenomenon presented in this work, would help form a complete picture of this complex system. 

\section{Conclusions}\label{sec:conclusion}

We report two new epochs of HST/STIS coronagraphic imaging of HD 163296, and new UV observations of its jet with HST/STIS. We also report the results of a multiwavelength (optical, near-IR, radio) monitoring campaign from 2016-2018. Finally, we reprocessed archival HST/STIS coronagraphic imaging epoch taken in 1998 September and recalculated accretion measurements from the Br$\gamma$ line taken from 2003-2013.  We found:

\begin{itemize}
\item{Ansae features previously identified by \citet{grady2000} are a 4th ring in the outer disk, having a semi-major axis value of 3$\farcs$25 (330 au) and a minor axis offset of  0$\farcs$7 (64 au). Assuming the 4th ring is perfectly elliptical, the scale height of the dust is 64 au at a radial distance of 330 au.}

\item{We observed surface brightness variations in the 4th ring (3$\farcs$25, 330 au) of the system's disk across all three of the HST/STIS epochs, including large-scale, azimuthally asymmetric changes between the 2018 April and the 2018 July epochs. Our data demonstrate that disk illumination variations occur on $<$ 3 months timescales. Since large-scale, azimuthally asymmetric changes in illumination have been observed both for the innermost disk ring ($0 \farcs 66$, 67 au \citealt{rich2019}) and now the fourth disk ring (3$\farcs$25, 330 au; \citealt{wisniewski2008} and current work), we suggest that a similar mechanism is responsible for these phenomenon.}

\item{We do not detect the ejection of new HH-knots predicted to occur in 2018 \citep{ellerbroek2014}. We also do not detect any of the HH-knots previously launched from HD 163296. These older knots could be experiencing less shock-heating and have cooled sufficiently to not be observable with our HST/STIS observations. The last detection of the jet associated with HD 163296 was in UV observations from 2017 August 10.}
\item{We detected one clear dipper event and a second, potential dipper event in our optical light curve from 2018.}

\item{Using the time-scale and spatial extent of disk illumination changes we observe, we estimate the source of this shadowing resides within 0.5 au from the star, must be at least 0.08 au above the midplane of the disk, and has an azimuthal extent of 0.26 au.  We estimate the source of the dipper event reaches a scale height of 0.37 au above the midplane at 0.41 au, and has an azimuthal extent of 0.3 au.  We suggest these similarities could indicate the same (or similar) mechanisms are responsible for producing both dippers and variable ring illumination in the system.}
\end{itemize}

Based on observations with the NASA/ESA Hubble Space Telescope obtained at the Space Telescope Science
Institute, which is operated by the Association of Universities for Research in Astronomy, Incorporated, under NASA contract NAS5-26555. Support for Program number GO-15437 was provided through a grant from the STScI under NASA contract NAS5-26555. This work has been supported by funding from the NASA XRP program NNX17AF88G. The authors recognize and acknowledge the significant cultural role and reverence that the summit of Maunakea has always had within the indigenous Hawaiian community. We are most fortunate to have the opportunity to conduct observations from this mountain. We acknowledge with thanks the variable star observations from the AAVSO International Database contributed by observers worldwide and used in this research. We would also like to thank Dr. Hans Moritz G\"{u}enther for your feedback on this manuscript, especially feedback on the interpretation of the UV observations. We thank the anonymous referee for their feedback that helped to improve this paper.
The National Radio Astronomy Observatory is a facility of the National Science Foundation operated under cooperative agreement by Associated Universities, Inc.

\clearpage

\begin{table*}
\small
\centering
\renewcommand{\arraystretch}{1.0}
\caption{Near-IR accretion measurements using Br$\gamma$.}
\begin{tabular}{lccccc}
\hline \hline
{HJD} & {K$_{obs}$} & {EW(Br$\gamma$)$_{obs}$} & {EW(Br$\gamma$)$_{cs}$} & {L(Br$\gamma$)} & {log $M_{acc}$(Br$\gamma$)} \\
{} & {(mag)} & {\AA} & {\AA} & {$(10^{-3}$ $\Lsun$)} & {($\Msun$ yr$^{-1}$)} \\
\hline
2002 Mar 23 $^a$ & 4.5 $\pm$ 0.05 & -3.0  & -7.2 $\pm$ 1.2 & 1.8 $\pm$ 0.3 & -6.73 $\pm$ 0.35 \\
2002 Jul 18 $^b$ & 4.5 $\pm$ 0.05 & -3.1 & -7.3 $\pm$ 1.2 & 1.8 $\pm$ 0.3 & -6.72 $\pm$ 0.35 \\
2004 Jun 09 $^c$ & 4.8 $\pm$ 0.05 & -4.7 & -10.2 $\pm$ 1.2 & 1.9 $\pm$ 0.2 & -6.69 $\pm$ 0.34 \\
2005 Jul 06 $^b$ & 4.6 $\pm$ 0.05 & -4.2 & -8.8 $\pm$ 1.2 & 2.0 $\pm$ 0.3 & -6.67 $\pm$ 0.34 \\
2008 Mar $^d$ & 4.8 $\pm$ 0.05 & -4.3 & -9.8 $\pm$ 1.2 & 1.8 $\pm$ 0.2 & -6.71 $\pm$ 0.34 \\
2008 May 13 $^d$ & 4.8 $\pm$ 0.05 & -4.3 & -9.8 $\pm$ 1.2 & 1.8 $\pm$ 0.2 & -6.71 $\pm$ 0.34 \\
2009 Jul 15 $^e$ & 4.8 $\pm$ 0.05 & -3.2 &  -8.7 $\pm$ 1.2 & 1.6 $\pm$ 0.2 & -6.77 $\pm$ 0.35 \\
2011 Oct 12 $^f$ & 4.8 $\pm$ 0.05 & -4.2 &  -9.7 $\pm$ 1.2 & 1.8 $\pm$ 0.2 & -6.71 $\pm$ 0.34 \\
2011 Oct 14 $^f$ & 4.5 $\pm$ 0.05 & -3.3 & -7.5 $\pm$ 1.2 & 1.8 $\pm$ 0.3 & -6.70 $\pm$ 0.35 \\
2011 Oct 16 $^f$ & 4.3 $\pm$ 0.05 & -3.9 & -7.4 $\pm$ 1.2 & 2.2 $\pm$ 0.4 & -6.61 $\pm$ 0.34 \\
2012 Mar 24 $^f$ & 4.3 $\pm$ 0.05 & -3.7 & -7.2 $\pm$ 1.2 & 2.1 $\pm$ 0.4 & -6.62 $\pm$ 0.35 \\
2012 May 17 $^f$ & 4.2 $\pm$ 0.05 & -3.7 & -6.9 $\pm$ 1.2 & 2.2 $\pm$ 0.4 & -6.60 $\pm$ 0.35 \\
2012 Jul 05 $^g$ & 4.7 $\pm$ 0.05 & -4.3 & -9.3 $\pm$ 1.2 & 1.9 $\pm$ 0.2 & -6.68 $\pm$ 0.34 \\
2016 Apr 04 & 5.07 $\pm$ 0.11 & -4.22 & -11.28 $\pm$ 1.2 & 1.6 $\pm$ 0.2 & -6.77 $\pm$  0.35 \\
2016 May 04 & 5.11 $\pm$ 0.11 & -5.33 & -12.66 $\pm$ 1.2 & 1.8 $\pm$ 0.2 & -6.72 $\pm$ 0.34 \\
2016 Jun 09 & 5.05 $\pm$ 0.11 & -4.37 &  -11.33 $\pm$ 1.2 & 1.7 $\pm$ 0.2 & -6.76 $\pm$ 0.34 \\
2016 Aug 10 & 5.0 $\pm$ 0.11 & -3.56 & -10.23 $\pm$ 1.2 & 1.6 $\pm$ 0.2 & -6.79 $\pm$ 0.35 \\
2016 Sep 07 & 5.14 $\pm$ 0.11 & -4.18 & -11.76 $\pm$ 1.2 & 1.6 $\pm$ 0.2 & -6.78 $\pm$ 0.35 \\
2017 May 25 & 5.13 $\pm$ 0.11 & -4.49 & -11.99 $\pm$ 1.2 & 1.6 $\pm$ 0.2 & -6.77 $\pm$ 0.34 \\
2017 Sep 12 & 4.96 $\pm$ 0.11 & -4.21 & -10.63 $\pm$ 1.2 & 1.7 $\pm$ 0.2 & -6.75 $\pm$ 0.35 \\
2018 Apr 08 & 4.94 $\pm$ 0.11 & -5.76 & -12.04 $\pm$ 1.2 & 2.0 $\pm$ 0.2 & -6.67 $\pm$ 0.34 \\
2018 Apr 16 & 5.01 $\pm$ 0.11 & -5.23 & -11.91 $\pm$ 1.2 & 1.8 $\pm$ 0.2 & -6.71 $\pm$ 0.34 \\
2018 May 16 & 4.76 $\pm$ 0.11 & -3.97 & -9.31 $\pm$ 1.2 & 1.8 $\pm$ 0.2 & -6.72 $\pm$ 0.35 \\
2018 Jun 24 & 4.77 $\pm$ 0.11 & -5.17 & -10.54 $\pm$ 1.2 & 2.0 $\pm$ 0.2 & -6.65 $\pm$ 0.34 \\
2018 Aug 11 & 4.95 $\pm$ 0.11 & -3.35 & -9.69 $\pm$ 1.2 & 1.6 $\pm$ 0.2 & -6.79 $\pm$ 0.35 \\
2018 Sep 22 & 4.85 $\pm$ 0.11 & -4.58 & -10.37 $\pm$ 1.2 & 1.8 $\pm$ 0.2 & -6.70 $\pm$ 0.34 \\
\hline
\end{tabular}
\tablecomments{
      \small
      \begin{itemize}
      \item Original magnitude and equivalent width measurements was taken from $a$: \citealt{brittain2007}, $b$: \citealt{sitko2008}, $c$: \citealt{garcia2006}, $d$: \citealt{donehew2011}, $e$: \citealt{mendigutia2013}, $f$: \citealt{ellerbroek2014}. 
      \item The accretion values (between 2001 - 2013) originally presented in \citet{ellerbroek2014} were recalculated above to include updated distance and radius values for HD 163296, and Br$\gamma$ to accretion luminosity relation is taken from \citet{fairlamb2017}.
      \end{itemize}}
\label{tbl:accretion}
\end{table*}

\end{document}